%% file: main.tex
\documentclass{article}

\usepackage{spconf}

\usepackage{amsmath,amssymb}
\usepackage{graphicx}
\usepackage{hyperref}

\usepackage{times}
\usepackage{epsfig}

\usepackage{algorithm}
\usepackage{algorithmic}
\usepackage{enumerate}
\usepackage{multirow}
\usepackage{multicol}
\usepackage{arydshln}
\usepackage{booktabs}
\usepackage{float}
\usepackage{subcaption}

\usepackage{xcolor}
\usepackage{soul}

\usepackage[english]{babel}

\input{defs}

\def\x{{\mathbf x}}

\title{Transforming Harmonic Coefficients for 3D Splat Compression}
\name{Tam Thuc Do$^\dag$, Philip A. Chou$^\star$, Gene Cheung$^\dag$
\thanks{The work of G. Cheung was supported in part by the Natural Sciences and Engineering Research Council of Canada (NSERC) RGPIN-2025-06252.}}
\address{$^\dag$York University, Canada ~~~~~~ $^\star${\tt packet.media}}
\begin{document}
\ninept
\maketitle
\begin{abstract}
We address the problem of color attribute compression for 3D splats.  We show that all images generated by 3D splats are linear in the coefficients for each color channel, each spherical harmonic, and each splat, and we identify a basis for the space of all such images.  We identify an inner product for the coefficient space that induces the squared error loss on images. We show that orthonormalizing the coefficients with respect to this innner product before coding can yield over 2 dB gain.
\end{abstract}
\section{Introduction}
\label{sec:intro}
\input{Sections/01-Intro}

\section{Related work}
\input{Sections/02-Related-works}

\label{sec:related_works}

\section{Technical Framework}
\input{Sections/03-Technical}
\label{sec:technical_framework}

\section{Experiments}
\label{sec:print}

\input{Sections/04-Implementation-Details}

\input{Sections/05-Experiments}
\label{sec:experiments}

\section{Conclusion}
\input{Sections/06-Conclusion}
\label{sec:conclusion}

\clearpage
\begin{small}
\bibliographystyle{IEEEbib}
\bibliography{refs}
\end{small}

\end{document}

%% file: defs.tex
\def\0{{\mathbf 0}}
\def\1{{\mathbf 1}}

\def\c{{\mathbf c}}
\def\d{{\mathbf d}}

\def\o{{\mathbf o}}

\def\r{{\mathbf r}}

\def\v{{\mathbf v}}

\def\x{{\mathbf x}}

\def\T{{\mathbf T}}

\def\cC{{\mathcal C}}

\def\cF{{\mathcal F}}

\def\bmu{{\boldsymbol \mu}}

\def\bPhi{{\boldsymbol \Phi}}

\def\bSigma{{\boldsymbol \Sigma}}

%% file: Sections/01-Intro.tex
3D splats have become the representation of choice for 3D radiance fields and novel view synthesis.
There are many types of 3D splat primitives, including 3D Gaussians, voxels, B-splines, local neural functions, etc. \cite{westover1990volume, KerblKLD:23, Sun2024SVR, FrankeRFS:24, ThomasS:25, ZhouZYWW:25, zhou2026splat}.
In general, each 3D splat is given by geometric parameters (e.g., location, orientation, shape, scale, density)---to represent density as a function of 3D space---as well as color attribute parameters (e.g., spherical harmonics in each color channel)---to represent color as a function of direction on the sphere.
Since any given scene may contain millions of 3D splats, each with numerous geometric parameters as well as dozens of color attribute parameters, compression is an important issue.
Numerous papers have addressed compression of the 3D splat parameters \cite{BagdasarianKLBHEM:25}.
In this paper, we address compression of the color attribute parameters, assuming that the desired geometry parameters have already been determined and compressed, and have been made available to both encoder and decoder.

Normally, geometric and color attribute compression are done jointly.  
Nevertheless, given whatever geometric parameters are chosen in that joint process, our framework can be applied to near-optimal compression of the chosen color attribute parameters.
The number of bits pertaining to the color attributes is typically far greater than the number of bits pertaining to the geometry, as demonstrated in many works (e.g., \cite{WangZHFDBC:24}) including our own.  
Therefore, there is a need to address color attribute compression with some precision.
Our work is agnostic to the splat primitive (Gaussian, voxel, etc.) but we present our results using voxel splats.
For many works on 3D splat compression, the color attributes comprise the first $M=16$ spherical harmonics for each of three color channels, for each splat, totaling 48 coefficients per splat.
Generally, directly quantizing in spherical harmonic space is not efficient for two reasons.

First, nearest-neighbor quantization---quantization designed to minimize the expected squared error---in spherical harmonics space (whether scalar or vector quantization, e.g., \cite{WangZHFDBC:24, wang2025nsvq, navaneet2025compgs}), is sub-optimal for minimizing squared error in image space, which is the desired objective.
Second, spherical harmonics may have inter-harmonic correlation that practical factorized entropy models have a difficult time representing.

In our paper, we first show that any images that may be generated by 3D splats with a given geometry are {\em linear} in the color attribute parameters, and we identify the natural basis for the vector space of such images, whose coefficients are the color attribute parameters.
We then identify an inner product on the coefficient space, which induces a norm on the coefficient space, such that the squared norm of the difference between coefficient vectors is the metric of interest, namely the squared error between their associated images. 

This is the metric used as the loss function that we seek to minimize, both to fit parameters and to code at a given bit rate.
With the Gram matrix defined as the matrix of inner products of the basis functions with themselves, we  propose to transform the color attribute coefficients with the square root of the Gram matrix.
This transform is equivalent to changing the natural basis to an othornormal basis.  
We show that this orthonormalizing transform is energy-preserving in that the square error in the coefficient domain is identical to the squared error in the image domain.
Thus, quantizing the coefficients to minimize the squared quantization error in the coefficient domain also minimizes the squared quantization error in the image domain.

Additionally, we appeal to high-resolution quantization theory to show that such energy-preserving transforms are optimal for coefficients that are to be quantized using uniform scalar quantization with equal step sizes.
Noting that any energy-preserving transform followed by a unitary transform is also energy-preserving, we propose, of all such energy-preserving transforms, the one that most reduces the cross entropy of a fully factorized entropy model. 
That is, the energy-preserving transform followed by an optimal energy-compaction transform, namely the KLT.

Finally, we show how to approximate both the energy-preserving and energy-compaction transforms by a tensor product of energy-preserving and energy-compaction transforms, along  spatial (splat index), directional (spherical harmonic), and color component axes of the color attribute coefficients.
This paper focuses on the gain of the energy-preserving and energy-compaction transforms on the directional (spherical harmonic) coefficients.
We demonstrate gains over 2 dB gain can be achieved by combining energy-preserving (i.e., orthonormalizing) and energy-compacting (i.e., decorrelating) transforms.
We contend that the papers capable of realizing these gains likely use learnable transforms of the SH and color coefficients that minimize distortion in image space.  Though nonlinear transforms such as MLPs would suffice, most of the gain would likely be achieved with a linear transform.

\vspace{0.05in}
\noindent
\textbf{Notation:}
Vectors and matrices are written in bold lowercase and uppercase letters, respectively.
The $(i,j)$ element and the $j$-th \textit{row} of a matrix $\mathbf{A}$ are denoted by $A_{i,j}$ and $\mathbf{A}_{j}$, respectively.
The $i$-th element in the vector $\mathbf{a}$ is denoted by $a_{i}$.
Operator $\|\cdot\|_p$ denotes the $\ell_p$-norm. 

%% file: Sections/02-Related-works.tex
Neural Radiance Fields (NeRFs) \cite{mildenhall2021nerf} represent scenes as continuous volumetric functions, but their high rendering cost motivated more explicit representations such as voxels and point-based primitives. 3D Gaussian Splatting (3DGS) \cite{kerbl20233d} represents scenes using anisotropic Gaussians with geometric, opacity, and view-dependent color attributes, enabling efficient rasterization. Sparse Voxel Rasterization (SVR) \cite{sun2025sparse} follows a similar explicit volumetric formulation. This shift toward localized primitives also connects 3DGS compression to point-cloud attribute compression.

The large number of Gaussian primitives and their high-dimensional attributes make compression essential. Existing approaches mainly exploit pruning, vector quantization (VQ), spatial transforms, and entropy modeling. Pruning methods such as CompGS\cite{navaneet2025compgs} and Compact3DGS\cite{lee2024compact3d} reduce the number of splats, often combined with vector quantization, while SizeGS \cite{xie2024sizegs} explicitly controls the splat count under a bit budget. Our approach instead targets color compression independently of the number of splats, making it complementary to pruning.

VQ is widely used for Gaussian attribute compression. NSVQ-GS applies VQ to a high-dimensional Gaussian representation, while CompGS \cite{wang2025nsvq} combines VQ with pruning. Other methods use entropy-constrained VQ (ECVQ) \cite{WangZHFDBC:24}. However, these approaches select codewords using Euclidean distance in coefficient space. Such distance does not reflect rendering distortion: the same coefficient error can produce different image errors depending on a Gaussian's visibility, opacity, spatial support, and contribution to the rendered image. This motivates considering the image-space distortion induced by rendering rather than coefficient-space error.

Other methods exploit spatial or statistical redundancy. EntropyGS \cite{huang2025entropygs} uses entropy models for Gaussian coefficients, while RAHT-based approaches \cite{gallina2025raht} exploit spatial correlation. Learned approaches such as FCGS \cite{chen2025fcgs}, ContextGS \cite{wang2024contextgs}, and HAC \cite{chen2024hac} use contextual or latent representations for entropy coding, while RAHLE \cite{sridhara2025rahte} and Compact3DGS \cite{lee2024compact3d} combine multiple compression mechanisms across Gaussian attributes.

Overall, existing methods primarily reduce splat count, quantize attributes, exploit spatial redundancy, or improve entropy coding. In contrast, our work studies the \textbf{relationship between splat’s coefficients distortion and image-space distortion}, and \textbf{introduces an orthonormal energy-preserving transform that aligns quantization error with rendering error}. This makes the proposed formulation \textbf{complementary} to pruning, VQ, spatial transforms such as RAHT, and learned entropy-coding methods.

%% file: Sections/03-Technical.tex
\subsection{Linearity in Directional Color Representation}
Let us begin with the 6D plenoptic function, $f(x,y,z,\vartheta,\varphi,\lambda)$, representing the radiance at wavelength $\lambda$ as seen from a point $(x,y,z)$ in direction $(\vartheta,\varphi)$ \cite{AdelsonB:91}.\footnote{The 7D plenoptic function includes the time $t$, but this paper concerns static radiance fields only.}  Henceforth we denote $\x=(x,y,z)$ as a point in 3D space, and we denote $\d=(d_x,d_y,d_z)$ as a unit vector in direction $(\vartheta,\varphi)$, so that the plenoptic function is denoted $P(\x,\d,\lambda)$.  The plenoptic function can describe arbitrary views of a scene by letting $\x$ be the camera center of projection $\o$ and letting $\d$ be the direction of the ray from $\o$ through a given pixel.

An effective way to model a plenoptic function is to apply the volumetric rendering integral to a radiance field \cite{mildenhall2020nerf,kajiya1984raytracing},
\begin{align}
    f(\o,\d,\lambda)
    = \int_0^\infty T(t;\o,\d) \sigma(\r(t)) L(\r(t),\d,\lambda) dt,
    \label{eqn:volumetric_rendering}
\end{align}
where $\r(t)=\o+t\d$ for $t\geq0$ is the {\em ray} from point $\o$ in direction $\d$, $\sigma(\x)$ is the {\em density} at point $\x$, $L(\x,\d,\lambda)$ is the radiant intensity at point $\x$ in direction $-\d$ at wavelength $\lambda$, and
\begin{align}
    T(t;\o,\d)
    = \exp\left(-\int_0^t \sigma(\r(s)) ds\right) .
\end{align}
The integral \eqref{eqn:volumetric_rendering} can be approximated using a finite number of 3D splats $\{g_n\}_{n=1}^N$ as
\begin{align}
    f(\o,\d,\lambda)
    \approx \sum_{n=1}^N T_n(\o,\d) \alpha_n(\o,\d) L_n(\d,\lambda) ,
\end{align}
where, denoting $g_n(\o,\d)$ as the integral of the density of the 3D splat $g_n$ along the ray $\r(t)=\o+t\d$,
\begin{align}
    \alpha_n(\o,\d) = \left(1 - \exp(-g_n(\o,\d))\right)
    \label{eqn:alpha_n}
\end{align}
is the probability that there is at least one particle in splat $g_n$ along the ray,
and
\begin{align}
    T_n(\o,\d)
    = \prod_{i=1}^{n-1}(1-\alpha_i(\o,\d))
    = \exp\left(-\sum_{i=1}^{n-1} g_i(\o,\d)\right)
    \label{eqn:T_n}
\end{align}
is the probability that there are no particles in any of the splats $g_1,\ldots,g_{n-1}$ along the ray, where the splats are ordered by increasing distance from $\o$ along the ray.
Thus,
\begin{align}
    \phi_n^s(\o,\d) = T_n(\o,\d) \alpha_n(\o,\d)
    \label{eqn:phi-s_n}
\end{align}
is the probability that splat $g_n$ is the first splat along the ray to have a particle, and hence $0\leq\sum_n\phi_n^s(\o,\d) \leq 1$.  It can be seen that
\begin{align}
    f(\o,\d,\lambda)
    \approx \sum_{n=1}^N \phi_n^s(\o,\d) L_n(\d,\lambda)
\end{align}
is a convex combination of the $L_n(\d,\lambda)$.

Assume each $L_n(\d,\lambda)$ is in the tensor product of functions on the sphere spanned by basis functions $\{\phi_m^d(\d)\}_{m=1}^M$ and functions on the real line spanned by basis functions $\{\phi_l^c(\lambda)\}_{l=1}^L$.  That is, assume each $L_n(\d,\lambda)$ can be expressed as a linear combination,
\begin{align}
    L_n(\d,\lambda)
    = \sum_{m=1}^M\sum_{l=1}^L \phi_m^d(\d) \phi_l^c(\lambda) c_{nml} ,
\end{align}
for some coefficients $c_{nml}$.
(For example, let $\{\phi_m^d(\d)\}_{m=1}^M$ be the first $M=16$ spherical harmonics and let $\{\phi_l^c(\lambda)\}_{l=1}^L$  be the $L=3$ emission spectra for red, green, and blue phosphors.)
Then
\begin{align}
    f(\o,\d,\lambda)
     & \approx \sum_{n=1}^N\sum_{m=1}^M\sum_{l=1}^L \phi_n^s(\o,\d) \phi_m^d(\d) \phi_l^c(\lambda) c_{nml} \\
     & = \bPhi\c ,
\end{align}
where $\bPhi$ is a row vector of basis functions, and $\c$ is a column vector of coefficients, both of length $K=NML$, and indexed by $k=(n-1)ML+(m-1)L+l$.
The coefficient $c_{nml}$ for the basis function $\phi_{nml}(\o,\d,\lambda)=\phi_n^s(\o,\d) \phi_m^d(\d) \phi_l^c(\lambda)$ can be interpreted as the coefficient for splat $n$, spherical harmonic $m$, and color channel $l$.  
We call the bases $\{\phi_n^s(\o,\d)\}$, $\{\phi_n^d(\d)\}$, and $\{\phi_n^c(\lambda)\}$ respectively {\em spatial}, {\em directional}, and {\em color} bases (hence the superscripts).  The basis $\{\phi_{nml}(\o,\d,\lambda)\}$ is the tensor product of these.

\vspace{0.1in}
\noindent
\textbf{Remark}: This shows clearly for the first time that the plenoptic functions
representable by a given set of 3D splats are {\em linear} in the coefficients $c_{nml}$.  These coefficients may be considered parameters of the {\em attributes} of the splats.  Note, however, that the plenoptic functions are generally {\em nonlinear} in the parameters of the {\em geometry} of the splats (such as, in the case of 3D Gaussian splats, the means $\bmu_n$, covariances $\bSigma_n$, and densities $\rho_n$), since those parameters determine the basis functions $\phi_n^s(\o,\d)$ nonlinearly through \eqref{eqn:alpha_n}--\eqref{eqn:phi-s_n}.

\subsection{Energy preserving and compacted transform}
Let $\cF:\{(\o,\d,\lambda)\}\rightarrow\mathbb{R}$ be the space of all plenoptic functions, let $\cF^o\subset\cF$ be the subspace of all plenoptic functions in the color space spanned by $\{\phi_l^c(\lambda)\}$ (i.e., $f(\o,\d,\lambda) = \sum_l \phi_l^c(\lambda) f_l^o(\o,\d)$ for some $f_l^o(\o,\d)$), and let $\cF_{\bPhi}=\{\bPhi\c\}\subset\cF^o$ be the subspace of plenoptic functions representable as $\bPhi\c$.  In particular, if $f\in\cF_\bPhi$ then $f(\o,\d,\lambda) = \sum_l \phi_l^c(\lambda) f_l(\o,\d)$ with
\begin{align}
    f_l(\o,\d)
    = \sum_n\sum_m \phi_n^s(\o,\d)\phi_m^d(\d)c_{nml} .
\end{align}

The first problem is to learn the parameters $c_{nml}$ of a set of 3D splats representing a given plenoptic function, given a set of color images.  
The value of pixel $i$ of these images, in color channel $l$, equals the sample $f_l^o(\o_i,\d_i)$ of the plenoptic function for color channel $l$ corresponding to a ray $\r_i(t)=\o_i+t\d_i$ through pixel $i$ in direction $\d_i$ from the camera's center of projection $\o_i$.  
Therefore, the plenoptic function $f=\bPhi\c$ that best fits the images minimizes the squared error between images,
\begin{align}
    D(f^o,f)
    = \sum_l \sum_i \left(f_l^o(\o_i,\d_i)-f_l(\o_i,\d_i)\right)^2 .
    \label{eqn:quadratic_objective}
\end{align}
This is an ordinary linear least squares problem, which has a unique solution $f^*=\bPhi\c^*$.
Finding $\c^*$ is commonly done using back-propagation of the gradients with some variant of gradient descent, but as we have shown, this is a convex (indeed, quadratic) problem, which is relatively easy to solve.

The next problem is to communicate the parameters $\c^*$ by quantizing them as $\hat\c$ and then entropy coding them with bit rate $R(\hat\c)$.
When reproduced as $\hat f = \bPhi\hat\c$,
the relevant distortion is the squared error between images, $D(f^*,\hat f)$.
For practical reasons, quantization is constrained to be uniform scalar quantization of the coefficients, and entropy coding is constrained to be independent (but not identical) coding of each coefficient.
Under these constraints, it is not optimal to quantize and entropy code $\c^*$ directly.
Instead, we propose to apply two transforms to $\c^*$ before uniform scalar quantization and entropy coding.
First, we apply an energy-preserving transform $\bar\c^*=\T_{ep}\c^*$,
and then we apply an energy-compaction transform $\bar{\bar\c}^*=\T_{ec}\bar\c^*$, before uniformly scalar quantizing $\bar{\bar\c}^*$ as $\hat{\bar{\bar\c}}$ and independently entropy coding the coefficients.
The quantized coefficients $\hat\c$ are recovered as $\hat\c=\T_{ec}^{-1}\T_{ep}^{-1}\hat{\bar{\bar\c}}$.

The rate-distortion benefits of these transforms are presented in the next section.  Here, we argue briefly why they are needed, and then show how to obtain them.

By $\bar\c^*=\T_{ep}\c^*$ being energy-preserving, we mean that the squared error in the coefficient domain, $||\hat{\bar\c}-\bar\c^*||^2$, is equal to the squared error in the image domain, $D(f^*,\hat f)$.  
Energy-preserving transforms are asymptotically optimal according to high-resolution quantization theory, assuming their coefficients are uniformly scalar quantized and optimally entropy coded \cite{grayneuhoff1998quantization}.  
To see this, assume that the quantization cells are sufficiently small, so that the probability density of $\bar\c$ over each quantization cell is roughly constant.  Then, regardless of the shape of each quantization cell, as long as it has the same volume, the following will not vary: the probability density over the cell, the probability $p$ of the cell, and the optimal number of bits $-\log p$ used to code the cell.  It can be shown that among uniform scalar quantizers of $\bar\c$ whose cells have the same volume, the one that minimizes $E||\hat{\bar\c}-\bar\c||^2$ has equal coefficient stepsizes, i.e., has cubic quantization cells.  
Since $\T_{ep}$ is energy-preserving, this quantizer also minimizes the expected distortion $ED(f^*,\hat f)$ for the given bit rate.  
Moreover, if any other transform were used, say $\c'=\T'\T_{ep}\c$, followed by uniform scalar quantization, this will result in quantization cells $\cC'$ whose inverse images $\bar\cC=\T'^{-1}\cC$ are {\em not} cubic or do not have the same volume (unless $\T$ is unitary), and hence would not be optimal.

The benefit of following $\T_{ep}$ with a unitary energy-compaction transform $\T_{ec}$ comes from the practical requirement of independent entropy coding of each coefficient.  
Energy compaction is measured as the difference between geometric and arithmetic means of the energy of each coefficient.  
When the coefficients are Gaussian, for high bit rates, this is the same as the coding gain for independently entropy-coded coefficients.  
It is a classic result that for Gaussian sources, the KLT has the optimal energy compaction or coding gain \cite{gersho1992vq}.  
In practice also, the KLT provides substantial improvements in the bit rate.  
Further, as it is unitary, it does not alter the high-rate optimality of the quantization cell shape provided by $\T_{ep}$.  
Thus, we use the KLT for $\T_{ec}$.

\subsection{Orthonormalization by induced Gram matrix}
At last, we show how to obtain $\T_{ep}$.  The key to defining an energy-preserving transform is defining an inner product that induces a norm whose square coincides with the distortion measure of interest, \eqref{eqn:quadratic_objective}.  Specifically, for $f,g\in\cF^o$, and a measure $\mu$ on $\{(\o,\d,\lambda)\}$, define the inner product
\begin{align}
    \langle f,g\rangle
    & = \int\int\int f(\o,\d,\lambda) g(\o,\d,\lambda) d\mu(\o,\d,\lambda) .
    \label{eqn:inner_product_0}
\end{align}
Choosing for $\mu$ a separable measure $d\mu(\o,\d,\lambda)=d\mu(\o,\d)d\mu(\lambda)$, and choosing for $d\mu(\o,\d)$ the counting measure on the points $\{(\o_i,\d_i)\}$, we have
\begin{align}
    \langle f,g\rangle
    & \! = \! \sum_i \!\! \int \!\!
        \left(\!\sum_l\!\phi_l^c(\lambda)f_l(\o_i,\d_i)\right) \!\!\!
        \left(\!\sum_k\!\phi_k^c(\lambda)g_k(\o_i,\d_i)\right) \! d\mu(\lambda) \\
    & \! = \! \sum_i\sum_l\sum_k \!
        \left(\int \! \phi_l^c(\lambda)\phi_k^c(\lambda)d\mu(\lambda)\right) \!
        f_l(\o_i,\d_i) g_k(\o_i,\d_i) .
    \label{eqn:inner_product}
\end{align}
Assume orthonormality of the basis functions $\{\phi_l^c(\lambda)\}$,\footnote{In this paper, we will not challenge the design of the basis functions of the color space, but rather will assume that the squared error of the coefficients of this space is perceptually meaningful.} so that \eqref{eqn:inner_product} reduces to
\begin{align}
     \langle f,g\rangle
     = \sum_i\sum_l f_l(\o_i,\d_i) g_l(\o_i,\d_i) .
\end{align}
From this it can be seen that \eqref{eqn:quadratic_objective} can be expressed
\begin{align}
    D(f^o,f) = ||f^o-f||^2 = \langle f^o-f,f^o-f\rangle .
\end{align}

Armed now with an inner product that induces the distortion measure of interest, we can explicitly express the $\c^*$ minimizing the distortion \eqref{eqn:quadratic_objective} as
\begin{align}
    \c^*
    = \arg\min_{\c} ||f^o-\bPhi\c||^2
    = (\bPhi^\top\bPhi)^{-1} \bPhi^\top f^o ,
\end{align}
where $\bPhi^\top = [\langle\phi_k,\cdot\rangle]$ denotes the length-$K$ (where $K=NML$)
column vector of functionals that are inner products with the $K$ basis functions in $\bPhi$, $\bPhi^\top f^o$ denotes the length-$K$ column vector of these functionals applied to $f^o$, and therefore $\bPhi^\top\bPhi$ denotes the $K\times K$ matrix of inner products of the basis functions with themselves, i.e., the {\em Gram matrix}.

We define the energy-preserving transform $\T_{ep}$ as the square root of the Gram matrix,
\begin{align}
    \T_{ep} = (\bPhi^\top\bPhi)^{1/2} .
\end{align}
Using $\bar\c^*=\T_{ep}\c^*$ and $\hat{\bar\c}=\T_{ep}\hat\c$, we have for all $f^*,\hat f\in\cF_{\bPhi}$
\begin{align}
    D(f^*,\hat f)
    & = ||f^*-\hat f||^2 = \langle f^*-\hat f,f^*-\hat f\rangle \\
    & = \langle \Phi(\c^*-\hat\c),\Phi(\c^*-\hat\c)\rangle \\
    & = (\c^*-\hat\c)^\top\Phi^\top\Phi(\c^*-\hat\c) \\
    & = (\bar\c^*-\hat{\bar\c})^\top\T_{ep}^{-1}(\Phi^\top\Phi)\T_{ep}^{-1}(\bar\c^*-\hat{\bar\c}) \\
    & = ||\bar\c^*-\hat{\bar\c}||^2 ,
\end{align}
thus proving that $\T_{ep}$ is energy-preserving.

Unfortunately, $NML\times NML$ transforms $\T_{ep}$ and $\T_{ec}$ are quite large.  However, they may be approximated by tensor products of $N\times N$, $M\times M$, and $L\times L$ matrices.  In our work, we use
\begin{align}
    \T_{ep} = \T_{ep}^s\otimes\T_{ep}^d\otimes\T_{ep}^c
\end{align}
where $\T_{ep}^s=(\bPhi^{s\top}\bPhi^s)^{1/2}$, $\T_{ep}^d=(\bPhi^{d\top}\bPhi^d)^{1/2}$, and $\T_{ep}^c=(\bPhi^{c\top}\bPhi^c)^{1/2}$ are respectively $N\times N$ spatial, $M\times M$ directional, and $L\times L$ color energy-preserving transforms, such that
\begin{align}
    (\bPhi^\top\bPhi) \approx (\bPhi^{s\top}\bPhi^s)\otimes(\bPhi^{d\top}\bPhi^d)\otimes(\bPhi^{c\top}\bPhi^c) .
    \label{eqn:gram_matrix_tensor_product}
\end{align}

To see \eqref{eqn:gram_matrix_tensor_product}, let $\r=(\o,\d)$ represent the ray $\r(t)=\o+t\d$, so that $\phi_{nml}(\o,\d,\lambda)=\phi_n^s(\o,\d)\phi_m^d(\d)\phi_l^c(\lambda)$ may be written $\phi_{nml}(\r,\d,\lambda)=\phi_n^s(\r)\phi_m^d(\d)\phi_l^c(\lambda)$.  Then for $f=\phi_{nml}$ and $g=\phi_{n'm'l'}$ in \eqref{eqn:inner_product_0}, we have
\begin{align}
    \langle \phi_{nml},\phi_{n'm'l'}\rangle
    & = \int\int\int \phi_n^s(\r)\phi_m^d(\d)\phi_l^c(\lambda) \nonumber \\
    & \times \phi_{n'}^s(\r)\phi_{m'}^d(\d)\phi_{l'}^c(\lambda)
    d\mu(\r,\d,\lambda) .
    \label{eqn:inner_product_2}
\end{align}
Choosing for $\mu$ a separable measure $d\mu(\r,\d,\!\lambda)\!\!=\!\!d
\mu(\r)d\mu(\d)d\mu(\lambda)$ and choosing for $d\mu(\r)$ and $d\mu(\d)$ the counting measure on the points $\{\r_i\}$ and $\{\d_i\}$, \eqref{eqn:inner_product_2} factorizes into
\begin{align}
    & \left[\sum_i \! \phi_n^s(\r_i)\phi_{n'}^s(\r_i)\right]\!\!\!
    \left[\sum_i \! \phi_m^d(\d_i)\phi_{m'}^d(\d_i)\right]\!\!\!
    \left[\int \!\! \phi_l^c(\lambda)\phi_{l'}^c(\lambda)d\mu(\lambda)\right] \nonumber \\
    & = \langle\phi_n^s,\phi_{n'}^s\rangle
    \times \langle\phi_m^d,\phi_{m'}^d\rangle
    \times \langle\phi_l^c,\phi_{l'}^c\rangle .
    \label{eq:factored_gram_matrix}
\end{align}
Note that the approximation comes from choosing a product measure on the points $\{\r_i\}$ and $\{\d_i\}$ instead of a joint measure on the points $\{(\r_i,\d_i)\}$.  However, our product measure has the same marginals.

We likewise use for the energy-compaction transform
\begin{align}
    \T_{ec} = \T_{ec}^s\otimes\T_{ec}^d\otimes\T_{ec}^c
\end{align}
where $\T_{ec}^s$, $\T_{ec}^d$, and $\T_{ec}^c$ are respectively $N\times N$ spatial, $M\times M$ directional, and $L\times L$ KLTs of slices of the coefficients of $\bar\c^*=\T_{ep}\c^*$.
For some experimental results we may alternatively use a joint $ML\times ML$ directional-color KLT $\T_{ec}^{dc}$ instead of $\T_{ec}^d\otimes\T_{ec}^c$.

The experimental results of the next section focus on evaluating the {\em directional} energy-preserving and energy-compaction transforms on the spherical harmonic coefficients.  Spatial energy-preserving and energy-compaction transforms for 3D splat coding have been proposed and evaluated in other works \cite{gallina2025raht, xie2025mesongs, sridhara2025rahte, do2026unrolling}, and we do not transform the given color space (unless with joint directional-color transforms).

%% file: Sections/04-Implementation-Details.tex
\begin{figure*}[ht]
    \captionsetup[subfigure]{labelformat=empty}
    \hspace{-3em}
    \begin{subfigure}[b]{0.7\textwidth}
        \centering
        \begin{subfigure}[b]{0.5\textwidth}
            \centering
            \includegraphics[width=\linewidth, trim={1.4cm 0.1cm 0.1cm 0.1cm},clip]{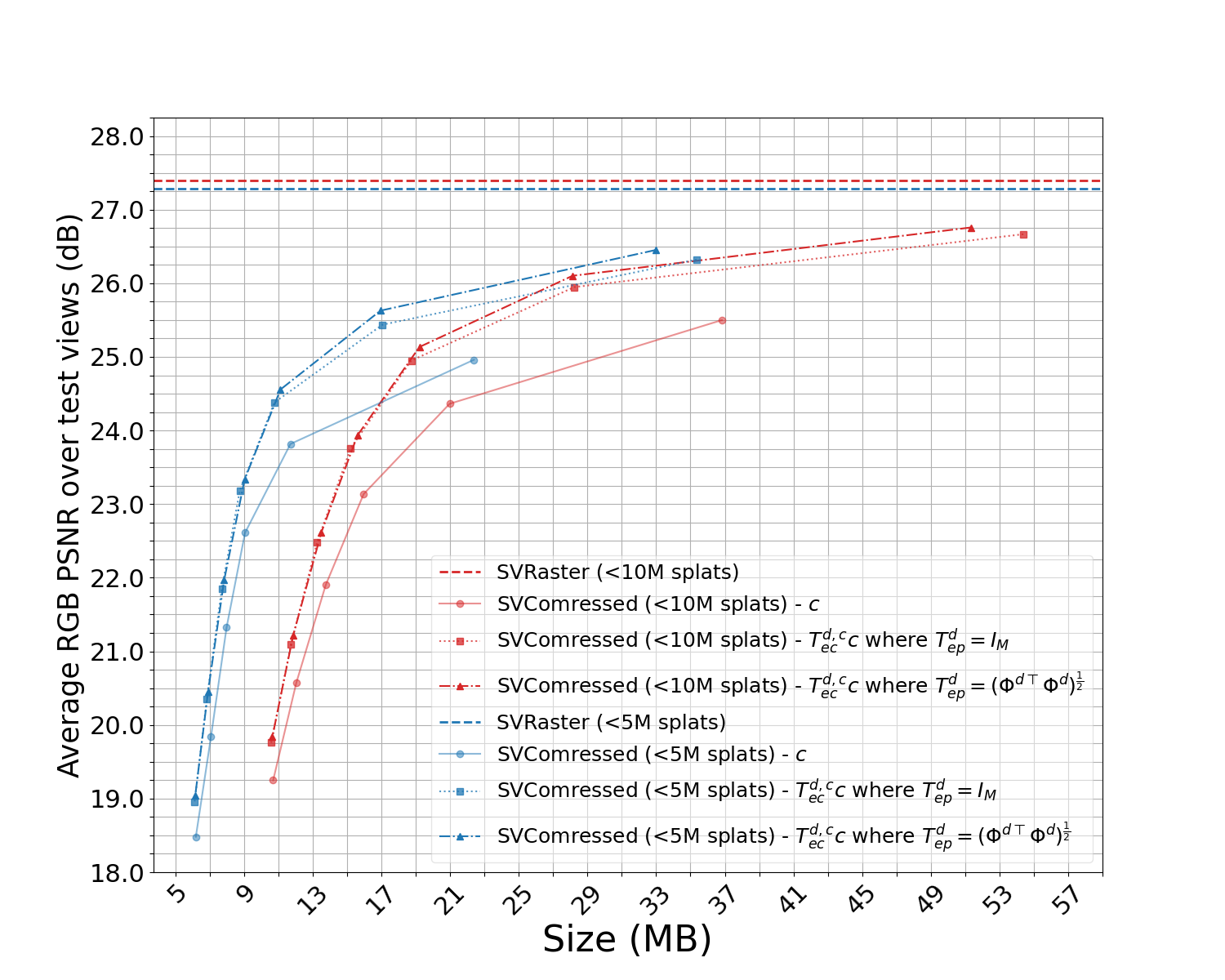}
            \caption{Mip Nerf 360}
        \end{subfigure}
        \hspace{-2.4em}
        \begin{subfigure}[b]{0.5\textwidth}
            \centering
            \includegraphics[width=\linewidth, trim={1.4cm 0.1cm 0.1cm 0.1cm},clip]{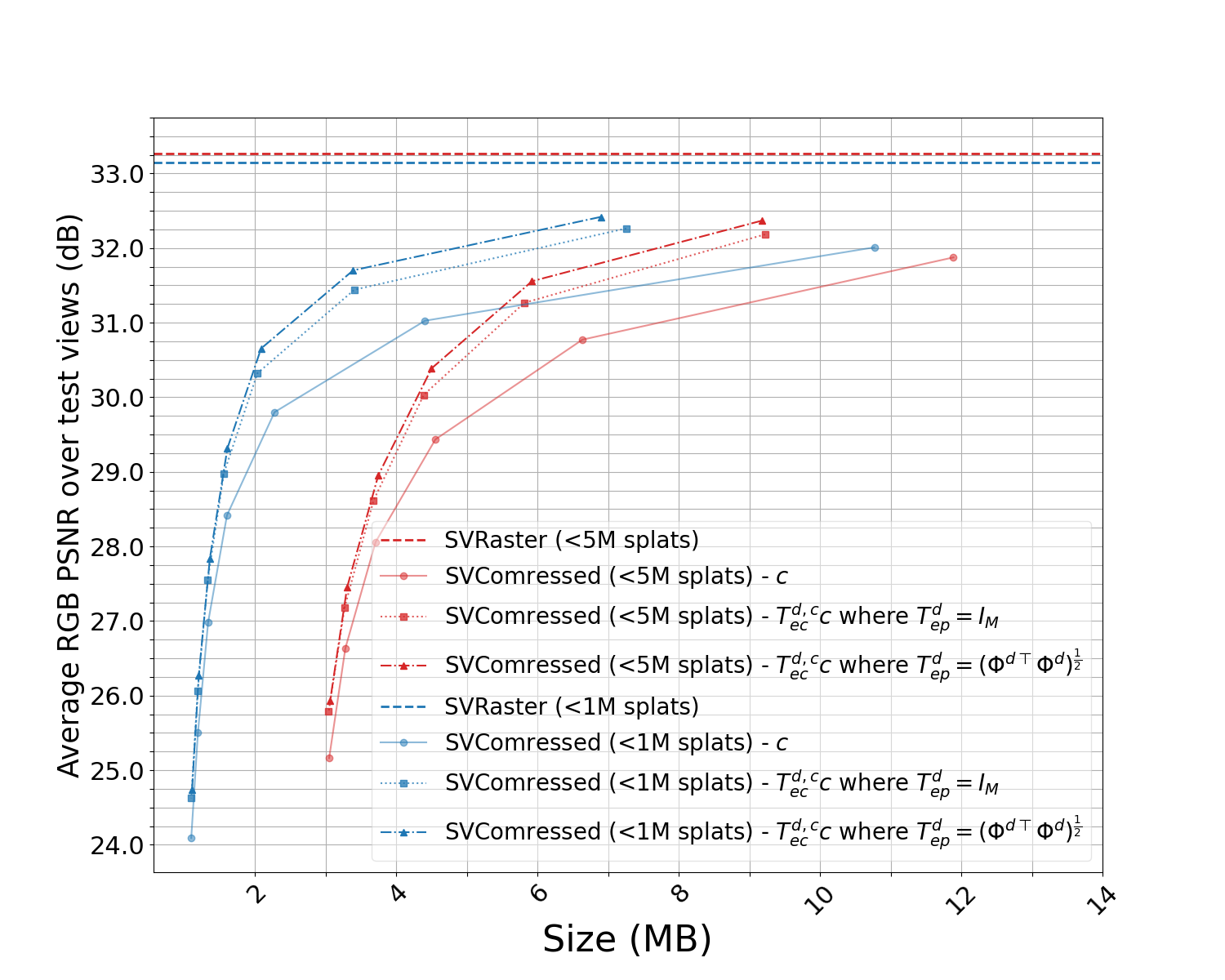}
            \caption{Nerf Synthetic}
        \end{subfigure}
        \caption{(a) Rate-Distortion Curves }
        \label{fig:RD-curvers-PartA}
    \end{subfigure}
    \hspace{-3em}
    \begin{subfigure}[b]{0.4\textwidth}
        \centering
        \begin{subfigure}[b]{0.48\textwidth}
            \centering
            \includegraphics[width=\textwidth]{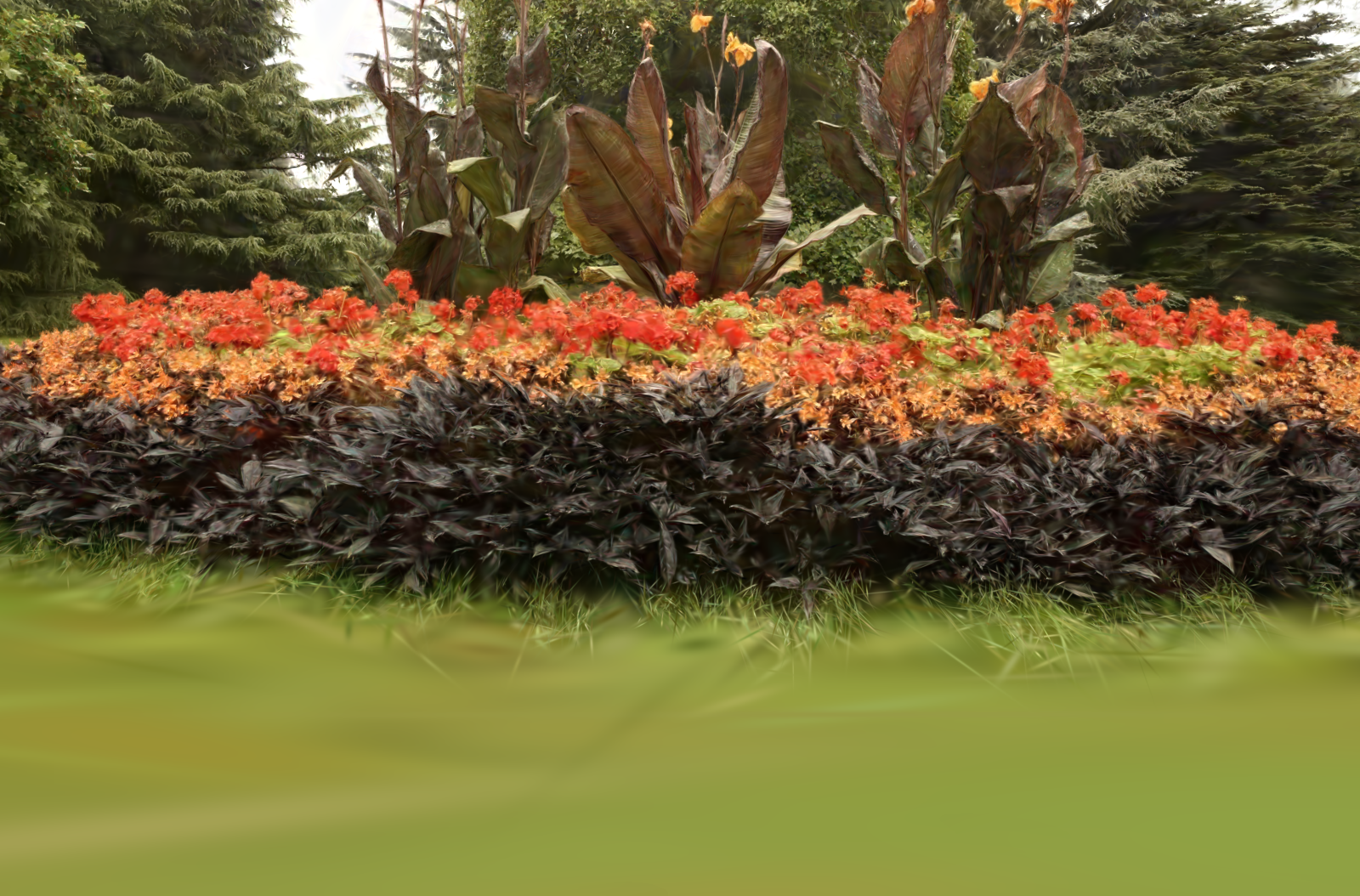}
            \caption{MesonGS (20.68 / 24.67)}
            \label{fig:sub1}
        \end{subfigure}
        \begin{subfigure}[b]{0.48\textwidth}
            \centering
            \includegraphics[width=\textwidth]{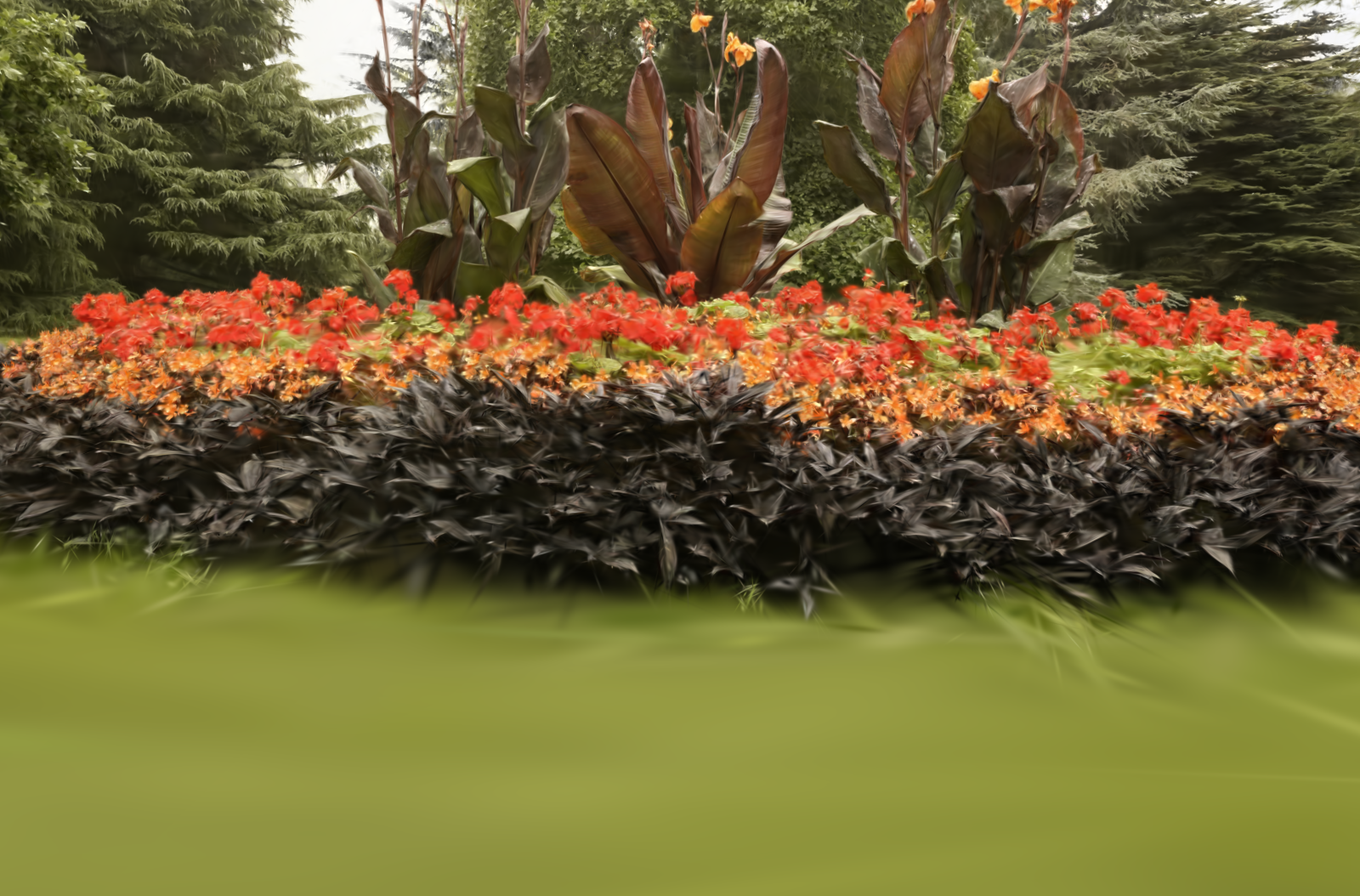}
            \caption{Compact3DGS (20.87 / 27.35)}
            \label{fig:sub2}
        \end{subfigure}\\
        \begin{subfigure}[b]{0.48\textwidth}
            \centering
            \includegraphics[width=\textwidth]{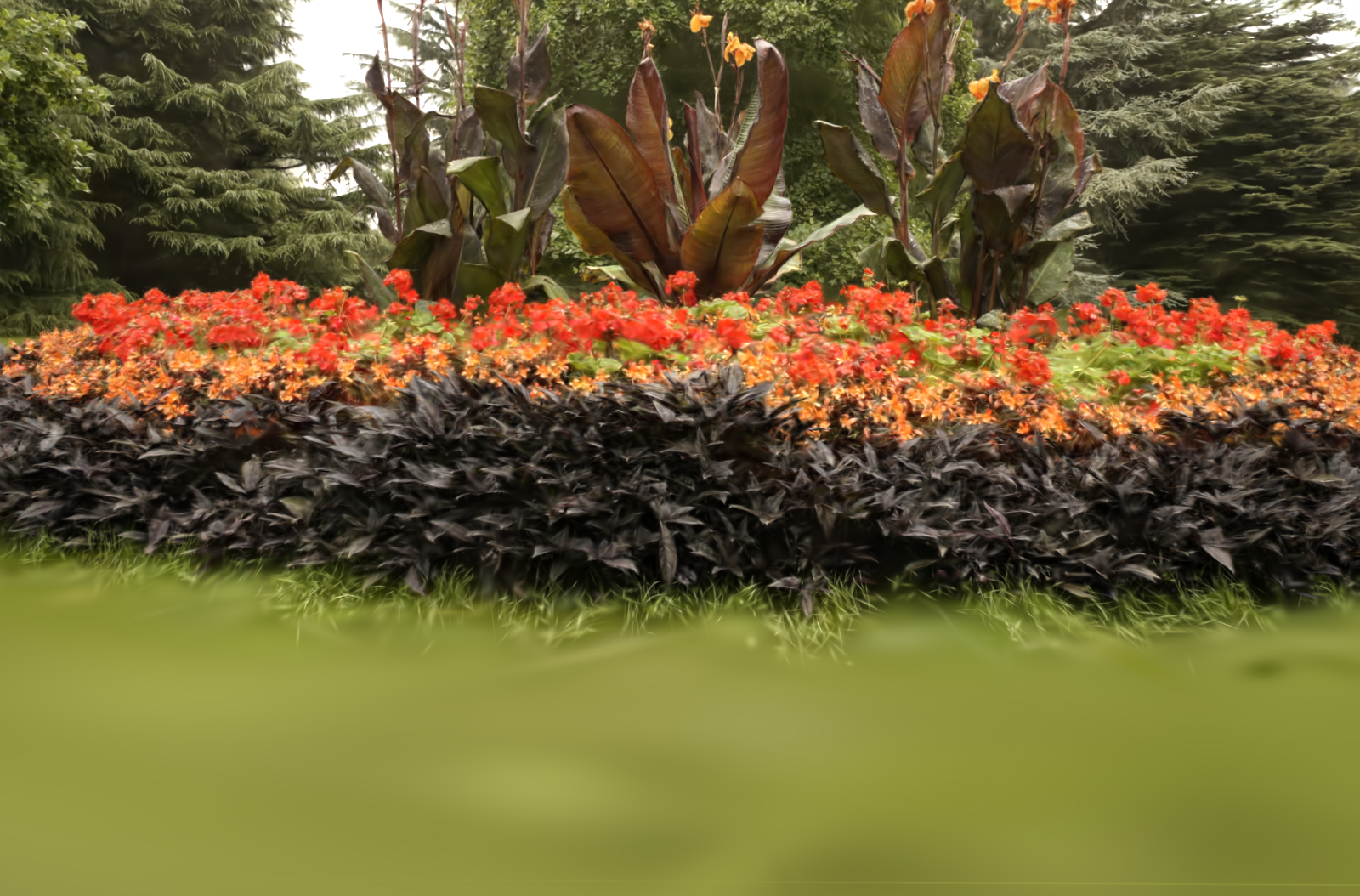}
            \caption{CompGS (21.11 / 26.24)}
            \label{fig:sub4}
        \end{subfigure}
        \begin{subfigure}[b]{0.48\textwidth}
            \centering
            \includegraphics[width=\textwidth]{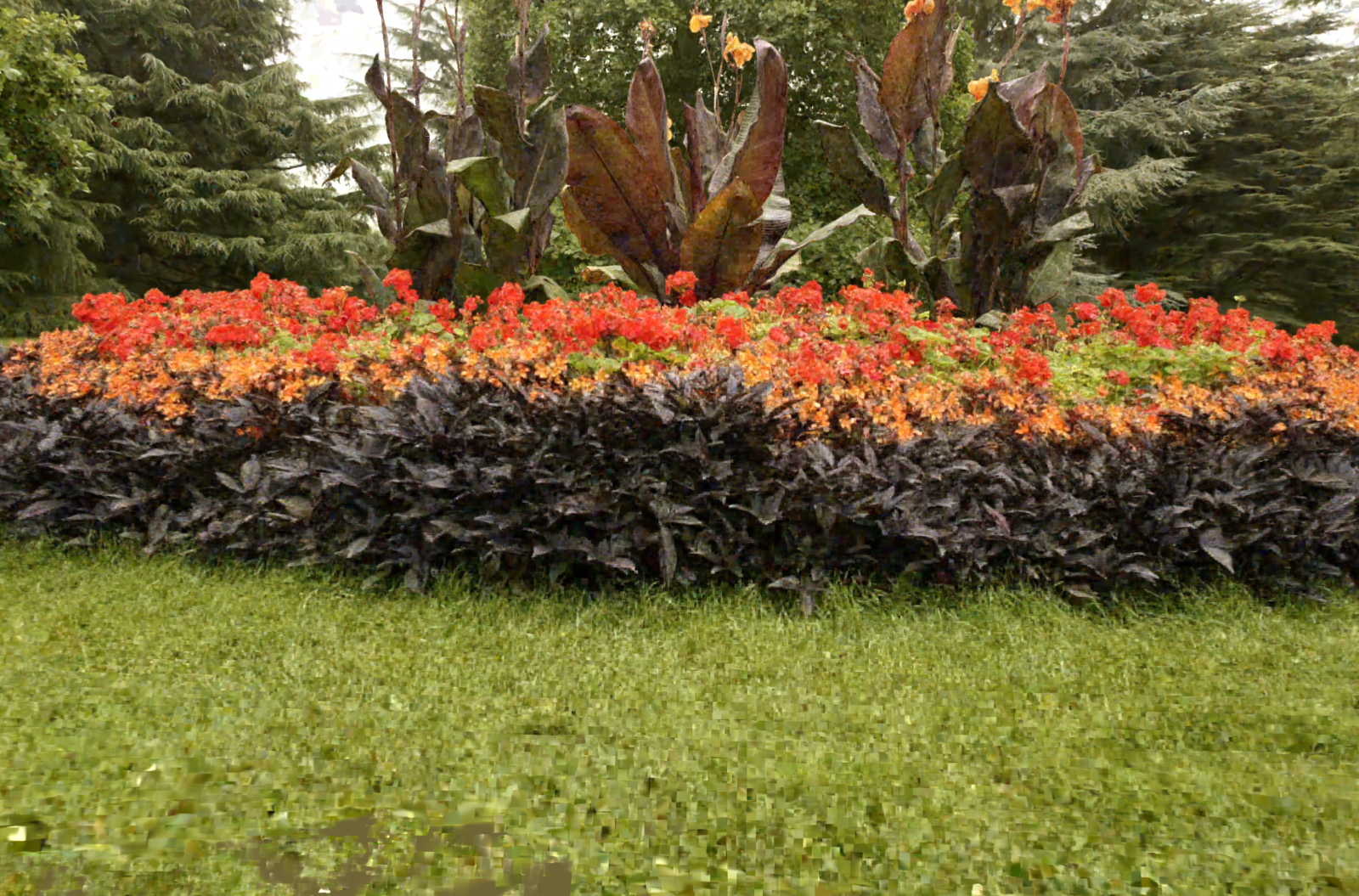}
            \caption{SVComp (20.84/21.19)}
            \label{fig:sub3}
        \end{subfigure}
        \vspace{1em}   
        \caption{(b) Rendered views of decoded bitsteams from different pipeline in (PSNR/MB) }
        \label{fig:rendered_views}
    \end{subfigure}
    \vspace{-2em}
\end{figure*}

%% file: Sections/05-Experiments.tex
\subsection{Implementation Details}
We now give details on the implementation of our encoding pipeline. As mentioned, we use Sparse Voxel Splats \cite{Sun2024SVR} as our splat representation due to their simplicity in parameterization of the density field $\sigma(\x)$.
For each ray-intersected voxels, a trilinear interpolation is applied using the eight parameters $\v_{g_n} \in \mathbb{R}^{2\times 2\times 2} $ 
attached to the voxel's corners to calculate the raw density, then an exponential-linear activation function is applied to map it to an non-negative volume density $g_n(\o,\d)$.
These voxel's corners and their parameters are considered as 16-bits point cloud attributes in our pipeline. 

Inherently, such density field representation has implicitly a degree of correlation locally, unlike Gaussian Splats (e.g., orientation, shape, scale), which is desirable by various developed spatial volumetric transforms (e.g., RAHT \cite{QueirozC:16, ChouKK:20, ThucTIP2026}). For the sake of simplicity, we decided to use RAHT as the base spatial transform operation $\T_{ec}^s$, and we refrained from using any deep learning based transforms that might distort the main points of our paper, even though these learned transforms would undoubtedly improve the coding performance over RAHT (\cite{IsikCHJT:22, do2024volumetric, zhan2025}).

Before encoding, the joint directional-color energy-compaction transform $\T_{ec}^{d,c}$ (the KLT) is estimated by scanning over all views that are to be considered in the distortion measurement to extract all possible directions for all voxels. Specifically, the matrix $\bPhi^{d\top}\bPhi^d$ of size $M \times M$ is calculated (with $M=16$) as in eq.\eqref{eq:factored_gram_matrix}, followed by its square-root $\T_{ep}^d = (\bPhi^{d\top}\bPhi^d)^{\frac{1}{2}}$. Then, the KLT is estimated using SVD on the matrix $\Sigma$ of size $ML\times ML$, 
\begin{equation}
    \Sigma[m_i, l_{i'}, m_j, l_{j'}] = \sum_n (\T_{ep}^d)_{m_i} \times c_{n m_i l_{i'}} \times c_{n m_j l_{j'}} \times (\T_{ep}^d)_{m_j}^T 
\end{equation}
where $(\T_{ep}^d)_{m_i}$ is the $m_i$ column of $(\T_{ep}^d)$ and $c_{n m l}$ is the coefficient of splat $n$, spherical harmonic $m$, and color channel $l$. We also compute a version of the KLT $\T_{ec0}^{d,c}$ where $\T_{ep}^d$ in (30) is replaced by the identity matrix.

Our encoding pipeline for the Voxel Splats is summarized as follows. We first use the occupancy octtree to encode the voxel's locations, followed by RAHT to spatially transform the density attributes and also the transformed directional-colors attributes. Then, the uniform quantization is applied and the quantized bin indices are fed into Asymmetric Numeral Systems (ANS) Entropy coder \cite{bamler2022understanding} for actual bit-stream encoding. Note that all the side-information (e.g., the directional-color KLT, quantized bin size, etc) is taken into account as meta-header, stored along with the bit-stream. For all evaluationd, we \textit{decode the bit-stream to reconstruct the voxel splats model} and then render the test views for distortion evaluation, and we also do this for all other methods.

\subsection{Experiments Details}
We used two datasets, Mip-nerf-360 and Nerf Synthetic, for evaluation.  The voxel splats are learned as introduced in \cite{sun2025sparse}, with two configurations corresponding to a large and a small number of splats. Given the learned splats, we simply encode them without requiring any further training or fine-tuning steps. For this work, we fix the quantization step sizes for density attributes with minimal distortion of rendered test views and vary the quantization step size for the transformed directional color attributes.

We estimate the rate-distortion curves for no SH transform and transform with or without directional transform (by setting $\T_{ep}^d$ to identity matrix). We also include other methods on 3D Gaussian Splat compression to contrast our simple pipeline with complex learning-based pipeline. We use the standard PSNR, LPIPS [60], and SSIM distortion metrics to evaluate novel-view quality.

\setlength{\tabcolsep}{2pt}
\begin{small}
\begin{table}[h]
    \vspace{-2mm}
    \makebox[0.5\textwidth][c]{
        \begin{tabular}{c|cccc}
            \toprule
            &\multicolumn{4}{c}{Mip Nerf 360} \\[-1pt]
            \cmidrule(lr){2-5}
            Method &Size(MB) &PSNR($\uparrow$) &SSIM($\uparrow$) &LPIPS($\downarrow$) \\[-1pt]
            \midrule
            3DGS
            &~700 &27.45 &0.815 &0.237 \\
            Sparse Voxels
            &~1762 &27.39 &0.819 &0.188 \\
            Compact-3DGS
            &15.41 &24.95 &0.720 &0.342 \\
            Meson-C3
            &28.33 &25.96 &0.769 &0.266 \\
            CompGS
            &21.90 &27.08 &0.802 &0.241 \\
            \midrule
            SVComp ($<$10M)
            &28.12 &26.10 &0.7461 &0.290 \\
            
            SVComp ($<$5M)
            &16.94 &25.62 &0.7187 &0.323 \\
            
            \hline
        \end{tabular}
    }
    \caption{\textit{Our result along with other compression pipelines}}
    \vspace{-6mm}
    \label{tab:result_with_other}
\end{table}%
\end{small}

\subsection{Results}

As shown in fig.~\ref{fig:RD-curvers-PartA}, the use of directional-color transform KLT significantly improves the coding (at least 1dB gain) in all configurations for both datasets. This emphasizes the handling of redundancy in spherical-hamonic basis for color representation contribute noticeably to the final encoding results. The plots also show the impact of number of splats in final encoded bit-streams (about 1 dB additional gain), which can be easily configured in the splat learning step without any fine-tuning or re-learning step in the encoding. In summary, we gain 2 dB with minimal computational overhead.

In addition, as shown in fig.~\ref{fig:rendered_views}, the use of sparse voxels splats allows us to avoid certain distortion artifacts entirely, while having nearly the same PSNR. This is also demonstrated in the table.~\ref{tab:result_with_other}, where sparse voxels show the best perceptual distortion. However, such fidelity requires significantly more storage, where despite of nearly $\times\frac{1}{64}$ size reduced, we still share similar compression performance as other pipelines. We leave the problem of learning a compact sparse voxels splats for our future works.

            
            

%% file: Sections/06-Conclusion.tex
We examine the problem of color attribute compression for 3D splats. We demonstrate the linearity in color representation of 3D splats under the spherical harmonic basis and provide an ortho-normalization operation, which contributes a significant coding gain to the compression pipeline. 